# Hole Transfer Dynamics in Thin Films of Mixed-Dimensional Quasi-2D Perovskites

G. Ammirati[1], S. Turchini[1], F. Toschi[1], P. O'Keeffe[2], A. Paladini[2], F. Martelli[1], R. Khanfar[3], D. Takhellambam[3], S. Pescetelli[3], A. Agresti[3,*] and D. Catone[1,*]

[1]*Istituto di Struttura della Materia - CNR (ISM-CNR), EuroFEL Support Laboratory (EFSL), Via del Fosso del Cavaliere 100, 00133, Rome, Italy.*
[2]*Istituto di Struttura della Materia - CNR (ISM-CNR), EuroFEL Support Laboratory (EFSL), Monterotondo Scalo 00015, Italy.*
[3]*Centre for Hybrid and Organic Solar Energy (CHOSE), Department of Electronic Engineering, Tor Vergata University of Rome, Via del Politecnico 1, 00133, Rome, Italy.*

*Corresponding authors: daniele.catone@cnr.it; antonio.agresti@uniroma2.it.

# Abstract

The understanding of charge transfer processes in mixed-dimensional quasi-2D perovskites is crucial for their application in high-performance photovoltaic devices. In this work, we investigate the link between charge transport dynamics and morphology in a thin film of quasi-2D perovskites ($PEA_2MA_{n-1}Pb_nI_{3n+1}$), grown with a distinct dimensionality gradient, where the n=1 phase is concentrated near the substrate and phases with higher dimensionality progressively increase in concentration toward the surface. By selectively exciting the n=4 phase, we observe efficient hole transfer to the n=2 and n=3 phases occurring within few tens of picoseconds after excitation. In contrast, the n=1 phase acts as a hole-blocking layer, limiting the overall charge transport efficiency. These results emphasize the critical importance of minimizing or eliminating the n=1 layer to enhance charge carrier separation and transport, offering valuable insights into the optimization of quasi-2D perovskite-based solar cells.

# Introduction

Halide perovskites possess remarkable optical and electrical properties, such as high absorption coefficients, high carrier mobility, long carrier diffusion lengths, and excellent defect tolerance[1–3]. These attributes have led to their widespread use in devices such as perovskite solar cells[4] (PSCs), light-emitting diodes[5], detectors[6], and lasers[7]. Notably, the power conversion efficiency (PCE) of single-junction PSCs surpassed 26% as of 2023, an unprecedented achievement among thin-film photovoltaic technologies[8]. However, the commercialization of 3D perovskites faces significant challenges due to their instability under environmental conditions[9]. The soft ionic lattice and weak noncovalent bonds between organic cations, like methylammonium ($MA^+$) and formamidinium ($FA^+$), and the inorganic framework make these materials susceptible to degradation in the presence of water[10]. Addressing the operational durability of perovskite materials is crucial for their commercial viability. Researchers have explored various strategies, but the most promising approach is transitioning from 3D perovskites to lower-dimensional structures without compromising the solar cell performances. Larger organic cations in 2D perovskites create a layered structure, effectively preventing direct contact between oxygen and moisture with the inorganic framework.[11] This property enhances environmental stability and allows for tuning the intrinsic physical properties such as electronic bandgap and exciton binding energy[12–15], providing more options for device performance customization.

Recent advancements have highlighted the potential of PSCs fabricated with mixtures of quasi-2D perovskites of varying dimensionalities. These systems exhibit a unique phase gradient, where phases with lower dimensionality (e.g., n=1) predominantly form near the substrate, while higher-dimensional phases accumulate closer to the surface[16,17]. This compositional gradient facilitates an efficient charge separation, promoting a directional transfer of charge carriers across different phases. The layered architecture optimizes charge transport pathways by minimizing energy barriers and enhancing separation efficiencies, making these quasi-2D perovskite films a promising approach to further boost device performance and operational stability for photovoltaic applications[18–20].

At the interfaces between the 2D perovskite layers, charge and energy transfer can take place via both type-I and type-II band alignment. In a type-I heterojunction, both charge carriers migrate from regions with larger bandgaps (lower dimensionality) to those with smaller bandgaps (higher dimensionality), facilitating charge funneling. On the other hand, type-II heterojunctions promote the spatial separation of electrons and holes, leading opposite charges to transfer in opposite directions, making them highly suitable for photovoltaic applications. The precise band alignment in 2D perovskites remains a subject of debate, with studies reporting both type-I[21–25] and type-II alignments[26–30], influenced by the material's chemical composition. In this context, the lack of a systematic characterization, along with the morphological variations, the limited number of perovskite phases and the challenges in reproducing these materials, makes it difficult to compare different works and can lead to seemingly contradictory results. Femtosecond transient absorption spectroscopy (FTAS), combined with photoluminescence (PL), has proven to be highly effective in investigating charge transfer processes and identifying gradients in phase distribution within quasi-2D perovskite systems, offering valuable insights into the interplay between morphology and charge dynamics in such materials[17,31].

In this work, we investigate the link between charge transport dynamics and morphology of a thin film formed by a mixture of $PEA_2MA_{n-1}Pb_nI_{3n+1}$ using FTAS and PL. Specifically, we explore how the hole transfer among layers is influenced by the distribution and morphology of phases with different dimensionalities within the quasi-2D perovskite mixture, as well as the relative band alignment between these phases, providing valuable insights into the role of interfacial coupling in governing charge transfer processes. Our findings highlight the critical importance of optimizing the layered structure to enhance the charge transport and the overall performance of 2D perovskite-based devices.

# Methods

## Sample preparation

The preparation of PEA-based quasi-2D perovskite thin films on glass (obtained by considering n=2 in the general formula $PEA_2MA_{n-1}Pb_nI_{3n+1}$), involves the following synthesis process. First, a precursor solution containing PEAI, MAI, and $PbI_2$ was mixed in a stoichiometric ratio of 2:1:2, respectively. Specifically, PEAI (0.144 g), MAI (0.046 g), and $PbI_2$ (0.276 g) were dissolved in 1 mL of DMF and stirred overnight.

Next, the glass substrate was thoroughly cleaned using a sequence of steps: it was washed with dishwashing soap, then sonicated for 15 minutes in a 1% glass cleaning solution, 10 minutes in distilled water, 15 minutes in acetone, and 15 minutes in 2-propanol. This was followed by a 15-minute UV-ozone treatment.

After cleaning, the substrates were pre-heated at 100°C. Subsequently, 100 μL of the perovskite precursor solution was deposited statically at 3000 rpm for 20 seconds (with an acceleration of 1500 rpm/s) and then annealed at 100°C for 10 minutes inside a glovebox. The crystallinity of the film and the orientation of the quasi-2D perovskites were investigated by X-Ray Diffraction (XRD), by using a Rigaku SmartLab SE 1D Type diffractometer working in Bragg–Brentano geometry equipped with a Cu K$\alpha$ source and a D/teX Ultra250 detector.

## Femtosecond Transient Absorption Spectroscopy

The experimental setup for femtosecond transient absorption spectroscopy (FTAS) consists of an amplified femtosecond laser (CARBIDE – Light Conversion, 180 fs pulse duration, 10 kHz repetition rate, 5 W power centered at 1030 nm) that generates tunable pump pulses using an optical parametric amplifier (Orpheus – Light Conversion). The probe is a white light supercontinuum beam (500–950 nm/1.30-2.50eV) generated by focusing 10 μJ of the 1030 nm pulse into a YAG crystal of 3 mm. The pump and the probe beams are focused on the sample with a diameter of 120 and 70 μm, respectively. The delay time between the two is changed by modifying the optical path of the probe, resulting in an instrument response function (IRF) of approximately 170 fs. The method used for the photoexcited carrier density estimation is reported in **Section S1.1** of the Supporting Information (SI).

**Photoluminescence**

Photoluminescence (PL) was measured using a 405 nm continuum wave laser of the MatchBox-2 series. The exciting beam was focused on the sample with a spot of 100 µm diameter, with an irradiance of 100 mW/cm$^2$. The PL signal was detected with a Peltier-cooled charge-coupled device after having been dispersed in a 30-cm-long monochromator with a 1200 grooves/mm grating.

## Results and discussion

The structural and optical properties of the $PEA_2MAPb_2I_7$ (PEA n=2) perovskite thin film were characterized using XRD, absorbance and PL spectroscopy, as reported in **Figure 1**. XRD analysis (**Figure 1a**) revealed diffraction peaks at 14.2°, 28.5°, and 43.3°, corresponding to the (111), (202), and (313) planes of the layered perovskite structure, respectively[32]. Notably, the peak at 43.4° is a clear indicator of the n=4 phase. Additional peaks below 14° were also observed, and were attributed to the (020), (040), and (060) planes. These peaks are associated with lower-n phases[33], suggesting a residual parallel crystal orientation. However, the very low intensity of these diffraction peaks confirms that the hot-casting deposition method employed successfully produced vertically-oriented quasi-2D perovskites, as schematically illustrated in the inset of **Figure 1a**.

The absorbance spectrum, reported in **Figure 1b**, revealed the presence of sharp excitonic peaks, at energy varying from 2.4 eV (n=1) to 1.9 eV (n=4) as expected for the different dimensionalities of the quasi-2D perovskite layers[16,27]. The PL spectrum exhibits Stokes-shifted peaks corresponding to the excitonic absorption features of the n=1 to n=4 phases of the $PEA_2MA_{n-1}Pb_nI_{3n+1}$ perovskite mixture.

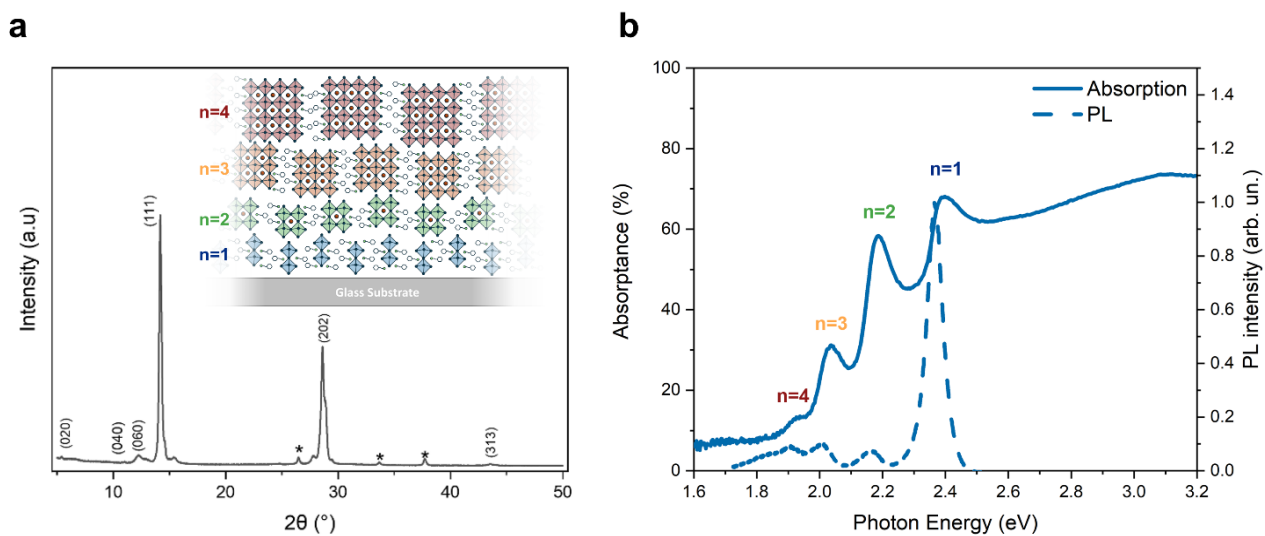

**Figure 1:** a) XRD patterns of PEA n=2 (PEA$_2$MAPb$_2$I$_7$) perovskite films, where * represents the FTO diffraction peaks. In the inset a scheme that shows the vertical orientation and distribution of the phases with different dimensionality within the thin film (from n=1 near the substrate to n=4 near the surface). The scheme is not representative of the relative abundance of each phase. b) Absorptance (full line, left axis) and PL (dashed line, right axis) spectra of the thin film of PEA$_2$MA$_{n−1}$Pb$_n$I$_{3n+1}$ showing excitonic peaks assigned to the n=1,2,3 and 4 phases as the energy decreases, respectively.

**Figure 2** presents the transient absorption (TA) spectra acquired at a time delay of 1 ps, following excitation of the thin film with a pump photon energy of 3.10 eV and a fluence of 6.2 μJ/cm² (the complete transient maps are reported in **Figure S1** of the SI). The TA spectra show the typical derivative-like line shape of quasi-2D perovskites, induced by various dynamic processes under pump excitation, such as band gap renormalization, screening of the exciton binding energy, and changes in electronic occupation[15,34,35]. These modifications induce, in the energy region of the exciton absorption, both negative and positive transient signals, namely photobleaching (PB) and photoinduced absorption (PIA), respectively. Additionally, intensity, shape, and minimum energy of the PB are influenced by Pauli blocking and phase space-filling, resulting from changes in electronic occupation within the system. The intensity of the TA spectra reported in **Figure 2** was evaluated by normalizing them to the total area of the PB signals to accurately assess the spatial distribution of the quasi-2D perovskite phases with different dimensionalities. For this reason, the sample was alternately illuminated from opposite sides (as illustrated in the inset of **Figure 2**).[31,36]

In this context, the intense negative signals at the photon energy of 2.36, 2.17, 2.02, and 1.92 eV were assigned to the excitonic absorption bleaching of n=1, n=2, n=3 and n=4, respectively[15,16]. The different intensity shown by the PB signals collected from opposite sides directly reflect the variation in the spatial distribution of the 2D phases. Indeed, when the sample is excited from the glass side, we observe a significant increase of the PB signal of the n=1 phase, accompanied by a reduction of the bleaching intensity in phases with n>2. These findings are in perfect agreement with the literature[26], confirming that the sample exhibits a higher concentration of the phase with n=1 near the glass substrate, while phases with higher dimensionality are predominantly located closer to the surface of the film.

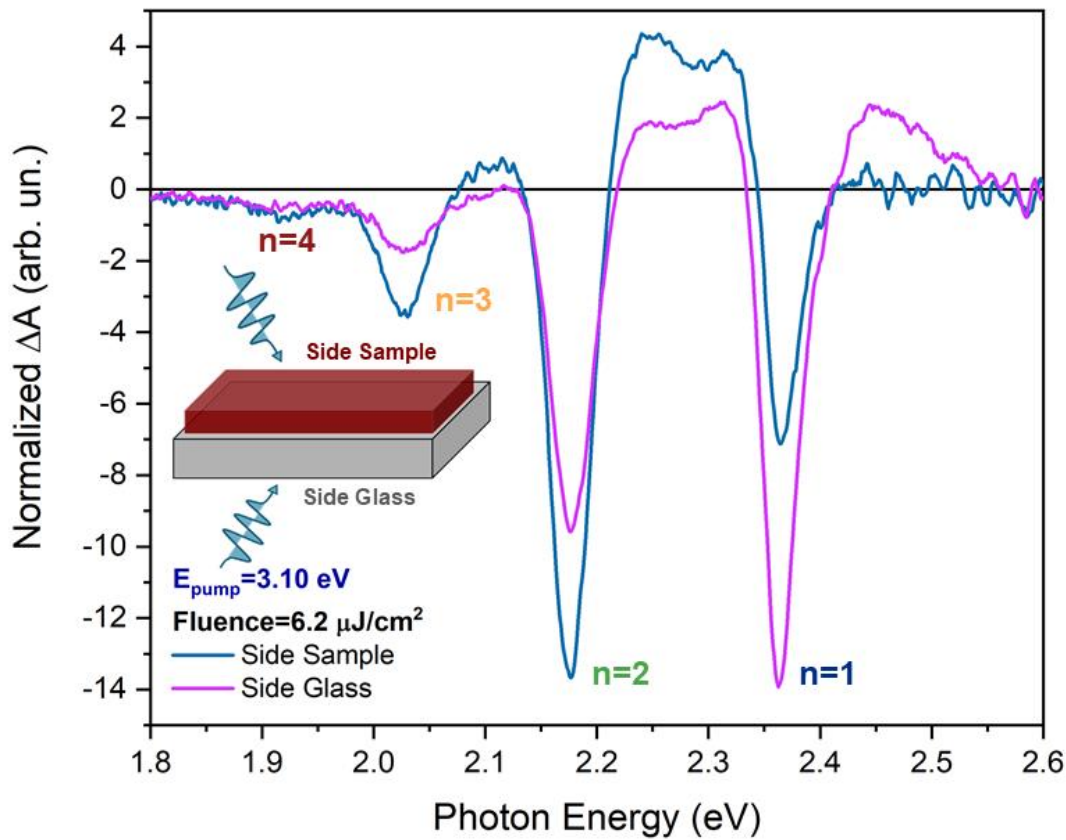

**Figure 2:** TA spectra at pump-probe delay of 1 ps following excitation of the PEA n=2 thin film with a pump photon energy of 3.10 eV and a fluence of 6.2 µJ/cm$^2$ (photoexcited carrier density=2.2·10$^{17}$ cm$^{-3}$). The blue TA spectra were collected by illuminating the sample from the surface side while the magenta TA spectra were collected by illuminating the sample from the substrate glass side (see inset). The intensity of the TA spectra was evaluated by normalizing them to the total area of the PB signals. The PB signals were assigned to the corresponding exciton resonances exhibited by the absorbance

spectrum and due to the n=1, 2, 3 and 4 phases. The relative intensities of the PB signals suggest preferential growth of the n=1 phase on the substrate glass side of the sample.

We can now discuss the temporal evolution of the TA signals. By analyzing the dynamics of these signals, we shed light on how the distribution and morphology of phases within the quasi-2D perovskite mixture, as well as the relative band alignment, influence the transfer of the excited charges.

**Figure 3a** shows the TA spectra acquired at selected time delays (t = 0.1, 0.5, 1, 5, 10, 50, 100, 150, 200, and 250 ps) with a pump at 1.90 eV and a fluence of 1550 µJ cm$^{-2}$ (the complete transient map is reported in **Figure S2** of the SI). The excitation energy selectively excites the n=4 phase. The TA spectra exhibit PB signals associated with the excitonic transitions of the n=2, n=3, and n=4 phases, but no PB signal is observed for the n=1 phase. Additionally, the PB signals show a decrease in intensity for the n=4 phase, accompanied by a gradual increase for the n=2 and n=3 phases during the first tens of picoseconds, as reported in **Figure 3b**.

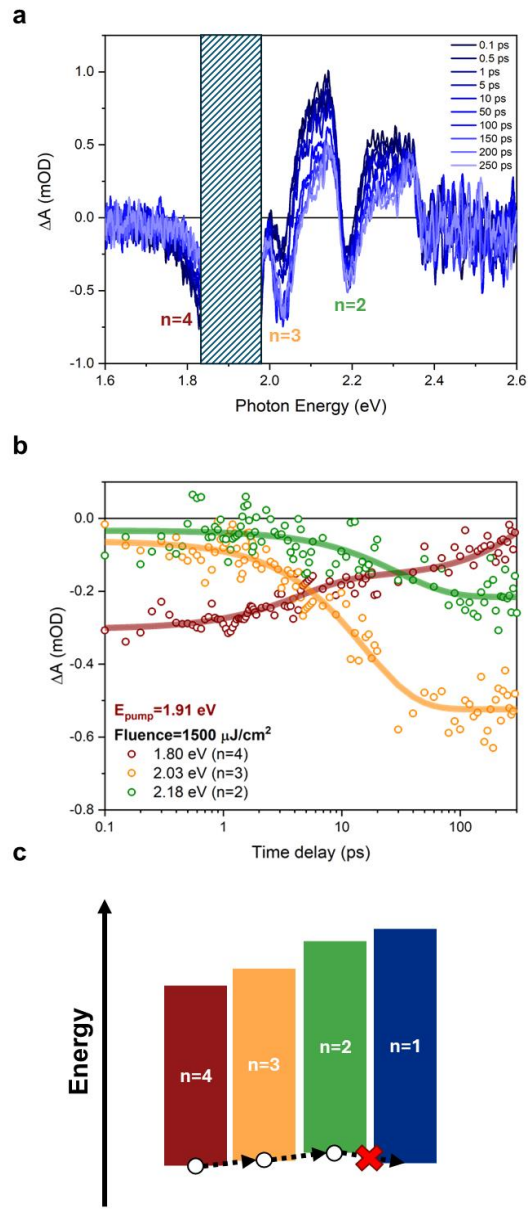

**Figure 3**: (a) TA spectra of the thin film of PEA n=2 acquired with a pump at 1.91 eV and 1550 µJ/cm$^2$ (n=1.9·10$^{19}$ cm$^{-3}$) and at selected pump-probe time delays (0.1, 0.5, 1, 5, 10, 50, 100, 150, 200, and 250 ps). The hatched box covers the region of the spectrum distorted by the scattered pump light. (b) Temporal dynamics of the n = 2, 3 and 4 PB signals as a function of time delays with corresponding exponential growth (n = 2 and 3) and biexponential decay (n=4) fits (note that signals are negative). This behavior suggests charge transfer from the photoexcited n=4 phase to the n=2 and 3 phases. (c) Schematization of the band alignment of the n = 1, 2, 3 and 4 phases of the PEA n=2 thin film deduced from the steady-state and time-resolved optical characterizations.

The dynamics of the PB signals for the n=4, n=3, and n=2 phases were obtained at the probe photon energies of 1.80, 2.03, and 2.18 eV, respectively. The results of the fitting procedure are reported in **Table 1**, and the fitting method is detailed in **Section S1.3** of the SI.

**Table 1.** The fit parameters extracted from the dynamics shown in Figure 3 were obtained using different models for each phase. For the n=2 and n=3 phases, a single rising ($\tau_r$) exponential function was applied, while for the n=4 phase, a bi-exponential ($\tau_1$ and $\tau_2$) decay function was employed.

| Probe Photon Energy (eV) | $\tau_r$ (ps) | |
|---|---|---|
| 2.18 (n=2) | 30±15 | |
| 2.03 (n=3) | 15±3 | |
| Probe Photon Energy (eV) | $\tau_1$ (ps) | $\tau_2$ (ps) |
| 1.80 (n=4) | 4.3±1.8 | 800±200 |

The appearance time of the PB signal for n=4 is limited by the IRF, and is thus consistent with the resonant excitation of the excitonic transition for this phase. Following this initial rise, the signal exhibits two distinct decay dynamics: a faster component ($\tau_1$), with a time constant of approximately 4 ps; a slower component ($\tau_2$), with a significantly longer time constant (on the order of hundreds of picoseconds), corresponding to radiative excitonic recombination. Conversely, n=3 and n=2 show only a rise time ($\tau_r$) of 15±3 ps and 30±15 ps, respectively. Hence, the slow rise of the PB signals for n=2 and n=3 is consistent with charge transfer originating from the excited n=4 phase that shows a decay component ($\tau_1$) attributed to processes that occur in the first picoseconds after excitation, namely bimolecular ultrafast recombination[37] and charge transfer. Considering these trends and the excitation of the phase with the lower excitonic transition, we deduce that the PB signals of n=2 and n=3 mainly arise from a hole transfer process. Thus, the longer time constant for n=2 with respect to n=3 suggests a sequential charge transfer, with charges progressively occupying states at lower energy, as sketched in **Figure 3c**.[31] Furthermore, the TA spectra does not exhibit any spectral feature associated with the n=1 exciton that should appear at about 2.35 eV, suggesting that hole transfer to this phase is inhibited. This behavior can be indicative of a band alignment that prevents the hole transfer from n=2 to n=1, suggesting a type I configuration between these two phases. On the other hand, the experimental evidence of the hole transfer from n=4 phase to n=3 and n=2, should suggest a type II configuration for these phases. This scheme only partially explains the PL spectrum. In fact, a type II configuration is expected to result in low PL efficiency due to charge separation, which is consistent with the observed low intensity of the peaks associated with the n=2, n=3, and n=4 phases [16,26]. Conversely, the high PL intensity of the n=1 phase indicates that this phase does not undergo an efficient charge transfer, as expected for a type I configuration with n=2, leading to a significantly lower PL intensity for n=1. For this reason, we cannot rule out that the lack of experimental evidence for hole transfer to the n=1 phase may arise from inefficient coupling between this phase and n=2, potentially caused by a lattice mismatch. The mismatch could introduce structural distortions at the interface, altering the electronic structure of the phases involved, reducing electronic overlap and leading to a band alignment that hinders efficient hole transfer. For example, changes in the Pb-I-Pb bond angle are known to significantly influence the bandgap in perovskites[29,38,39].

Additionally, the high abundance of low-dimensionality phases, due to the precursor molar ratios used (optimized for the n=2 phase) and evidenced by the relative intensity of the PB signals shown in **Figure 2**, results in a large layer thickness of the n=1 phase. Such a feature would explain an intense PL from the n=1 phase also in the presence of carrier transfer to n=2 layers as the low diffusion of the excitons strongly limits their possibility to reach the interface and effectively contribute to charge transfer.

This discussion highlights the fact that the topic still presents several open questions, requiring further theoretical and experimental investigations in the future to fully unravel the underlying mechanisms governing the charge transfer processes in quasi-2D perovskite-based materials.

In conclusion, a thin film of PEA-based quasi-2D perovskite was successfully grown, showing a controlled gradient of mixed-dimensional quasi-2D perovskites with the abundance of phases with higher dimensionality progressively increasing toward the surface of the film. Specifically, we demonstrated how the controlled mixture of quasi-2D perovskites induces an efficient hole transfer between phases with dimensionalities n≥2 that occurs in the first tens of picoseconds after excitation. In contrast, the n=1 phase acts as a hole-blocking layer, limiting overall charge transport efficiency. These findings underscore the importance of minimizing, possibly eliminating, the n=1 layer in solar cell designs to enhance charge carrier transport and separation throughout the active layer. Additionally, the observed spatial phase distribution and band alignment suggest that a p-i-n configuration is the most favorable architecture for these materials. This configuration leverages the dimensionality gradient to facilitate charge separation, directing holes toward the substrate side and electrons toward the surface, ultimately improving device performance. These results provide a robust framework for the design and optimization of quasi-2D perovskite-based solar cells, paving the way for achieving high efficiency and operational stability in next-generation photovoltaic devices.

## Acknowledgements

S.T., G.A., D.C., P.O.K. acknowledges for funding the European Union – NextGenerationEU, M4C2, within the PNRR project NFFA-DI, CUP B53C22004310006, IR0000015. D.C., S.T., D.C., P.O.K., A.P., F.T. and A.A. acknowledge financial support under the National Recovery and Resilience Plan (NRRP), Mission 4, Component 2, Investment 1.1, Call for tender No. 104 published on 2.2.2022 by the Italian Ministry of University and Research (MUR), funded by the European Union – NextGenerationEU – Effective Light management in 2D perOvskite absorbeRs for A disruptive tanDem phOtovoltaic technology (ELDORADO) – CUP E53D23001720001 - Grant Assignment Decree No. 957 adopted on 30/06/2023 by the Italian Ministry of Ministry of University and Research (MUR).

## Author Contributions

D.C. and G.A. conceived the article. G.A., D.C., P.O.K., F.T., and A.P. carried out transient and steady-state optical measurements. The data analysis was performed by G.A. with the supervision of D.C., S.T., and F.M.. R.K., D.T., S.P., and A.A. synthesized the materials and characterized them by X-ray diffraction. All experimental and theoretical findings were collaboratively discussed by the authors. The initial draft of the manuscript was prepared by D.C. and G.A., with all authors contributing to its writing, review, and finalization. S.T., D.C. and A.A. are responsible for the founding acquisition.

# Supporting Information

## S1. Femtosecond Transient Absorption Spectroscopy (FTAS)

### S1.1 Photoexcited carrier density estimation

**Equation S1** provides the formula to estimate the photoexcited carrier density ($n_0$) in FTAS

$$n_0 = \frac{F}{\hbar\omega} \cdot \frac{A}{d} \qquad \text{S1}$$

Where F is the pump fluence, $\hbar\omega$ is the pump photon energy, A is the absorptance of the sample and d is the sample thickness. **Table S1** reports the data used for the evaluation of the photoexcited carrier density.

| Photon Energy (eV) | Fluence (µJ/cm²) | A (%) | Thickness (nm) | $n_0$ (cm$^{-3}$) |
|---|---|---|---|---|
| 3.10 | 6.2 | 70 | 400 | $2.2 \cdot 10^{17}$ |
| 1.90 | 1550 | 15 | 400 | $1.9 \cdot 10^{19}$ |

**Table S2:** Carrier density estimated at the pump photon energy and fluence, along with the relative sample thickness and absorptance at the pump photon energy.

### S1.2 Transient Map of PEA-based perovskites at different pump conditions

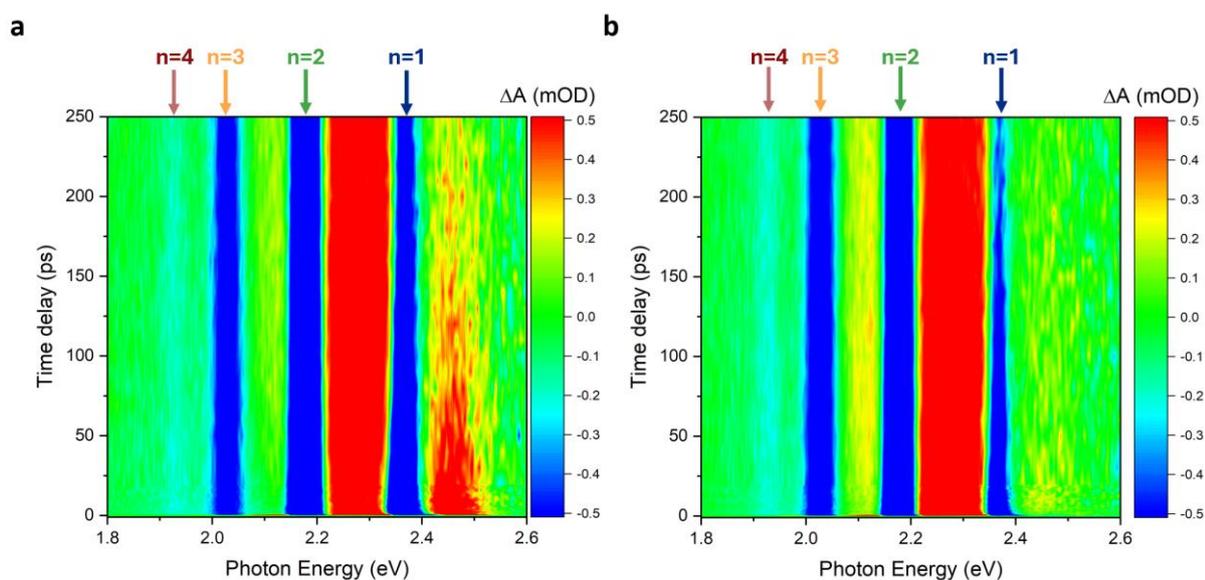

**S1:** TA maps of PEA n=2 thin film acquired with a pump photon energy of 3.10 eV and a fluence of 6.2 µJ/cm² (photoexcited carrier density=2.2·10¹⁷ cm⁻³). The TA maps were collected (**a**) by illuminating the sample from the surface side and (**b**) by illuminating the sample from the substrate glass side. The PB signals indicated by the arrow were assigned to the corresponding exciton resonances of the n=1, 2, 3 and 4 phases.

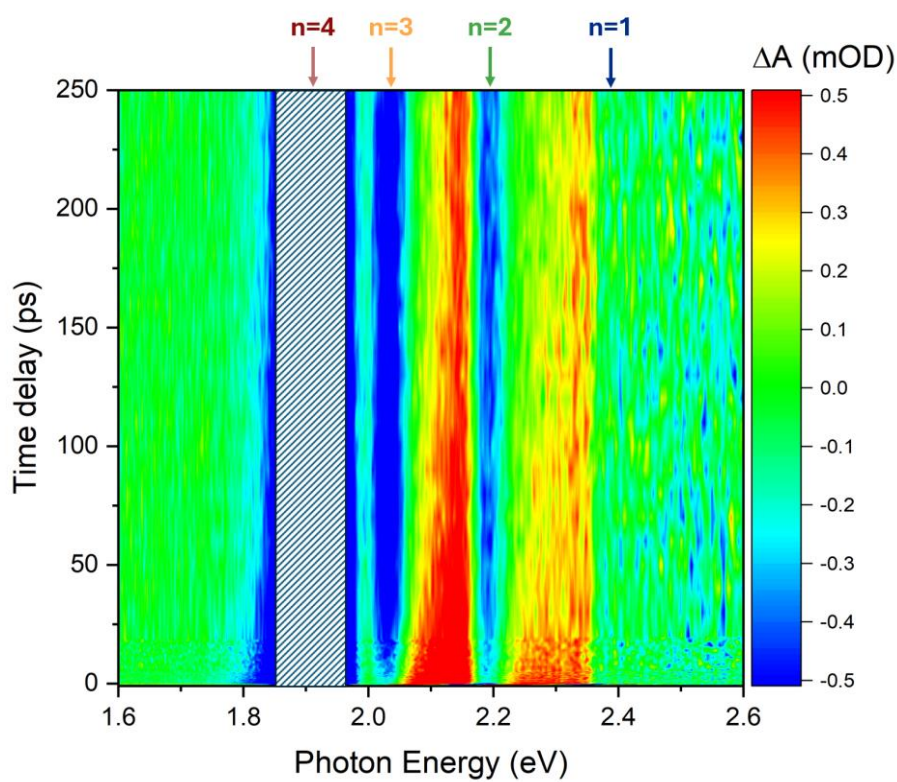

**Figure S2**: The TA map of the thin film of PEA n=2 thin film acquired with a pump at 1.91 eV and 1550 µJ/cm² (n=1.9·10¹⁹ cm⁻³ ) showing the evolution of the n = 2, 3 and 4 PB signals (n=1 signal is absent). The hatched box covers the region of the spectrum distorted by the scattered pump light. The TA spectra shown in **Figure 3b**, were extracted at selected time delays from the showed TA map.

## S1.3 Fitting Equation used for the PB temporal dynamics

**Equation S2** shows the fitting formula used for the dynamics of the n=4 phase reported in **Figure 3b**:

$$\Delta A_{n=4}(t) = \sum_{j=1}^{2} A_j e^{-(t-t_0)/\tau_j} \qquad \text{S2}$$

Where $A_j$ is the decay amplitude of the j-th component, $\tau_j$ is the j-th time decay, and $t_0$ is the fitting starting time (fixed at 0.5 ps).

**Equation S3** shows the fitting formula used for the rise dynamics of the n=2,3 reported in **Figure 3b**

$$\Delta A_r(t) = A_r \cdot \left(1 - e^{-\frac{t-t_0}{\tau_r}}\right) + A_l \qquad \text{S3}$$

Where $A_r$ is the rise amplitude, $\tau_r$ is rise time, and $t_0$ is the fitting starting time (fixed at 0.5 ps), and $A_l$ is the amplitude component that considers all the radiative and non-radiative recombinations taking place on a timescale longer with respect to our temporal window of 300 ps.

The time constants evaluated by these fitting procedures are reported in **Table 1** of the main text.